\documentclass[prd, twocolumn, 10pt]{revtex4-1}

\usepackage{bm}
\usepackage{epsfig}
\usepackage{amsmath}
\usepackage{latexsym}
\usepackage{slashed} 
\usepackage{mathrsfs, amsfonts, amssymb}

\newcommand{\be}{\begin{equation}}
\newcommand{\ee}{\end{equation}}
\newcommand{\bea}{\begin{eqnarray}}
\newcommand{\eea}{\end{eqnarray}}
\newcommand{\bma}{\begin{matrix}}
\newcommand{\ema}{\end{matrix}}
\newcommand{\nn}{\nonumber}
\newcommand{\bml}{\begin{mathletters}}
\newcommand{\eml}{\end{mathletters}}
\newcommand{\bes}{\begin{subequations}}
\newcommand{\ees}{\end{subequations}}

\newcommand{\bi}{\begin{itemize}}
\newcommand{\ei}{\end{itemize}}
\newcommand{\half}{\frac{1}{2}}

\newcommand{\tBar}{\bar{\theta}}

\begin{document}

\title{AdS$_5$ Supersymmetry in $N=1$ Superspace}
\author{Jonathan Bagger}
\email[]{bagger@jhu.edu} 
\affiliation{Department of Physics and Astronomy, Johns Hopkins University, 
3400 North Charles Street, Baltimore, MD 21218, U.S.A.}

\author{Chi Xiong}
\email[]{xiong@virginia.edu}
\affiliation{Department of Physics, University of Virginia, 
382 McCormick Road, P.O.\ Box 400714, Charlottesville, VA 22904-4714,
U.S.A.}

\date{\today}
\begin{abstract}
We use $N=1$ superspace to construct the supersymmetric matter couplings of vector and hyper multiplets in a five-dimensional anti-de Sitter spacetime background.  For hypermultiplets, we find that AdS$_5$ supersymmetry requires the scalar fields to lie on a hyper-K\"ahler manifold endowed with a certain type of holomorphic Killing vector.
\end{abstract}

\maketitle
\section{Introduction}

In four flat dimensions, supersymmetric nonlinear sigma models have a natural interpretation in terms of K\"ahler geometry.  The scalar fields are the coordinates of a K\"ahler manifold; the vector fields gauge its holomorphic isometries.  In five flat dimensions, supersymmetric theories enjoy a similar description. The scalar fields are the coordinates of a hyper-K\"ahler manifold, while the vector fields gauge its tri-holomorphic isometries \cite{Early}.

In this paper we extend this story to five-dimensional anti-de Sitter backgrounds.  We work in superspace to make the supersymmetry manifest.  However, five-dimensional superfields generally contain an infinite number of auxiliary fields \cite{5dss}.  To avoid this issue, we use what are essentially $N=1$ superfields \cite{MSS}--\cite{BX1}.  Our approach spoils manifest invariance under half the AdS$_5$ supersymmetries.  Nevertheless, it provides a simple and convenient superspace framework for supersymmetric theories in AdS$_5$.  It offers a natural way to construct bulk-plus-brane scenarios in which flat three-branes foliate a five-dimensional AdS$_5$ bulk.  

We illustrate our approach using vector and hyper multiplets, and construct AdS$_5$ versions of supersymmetric Yang Mills theory and the supersymmetric nonlinear sigma model.  In the latter case, we show that the scalar fields must parametrize a special type of hyper-K\"ahler manifold, one that admits a holomorphic Killing vector that obeys a certain condition on the target space.  

This paper is organized as follows. In Sect.\ II we review the formulation of five-dimensional supersymmetric nonlinear sigma models in $N=1$ superspace. In Sect.\ III we introduce warped superspace as an appropriate venue for constructing supersymmetric theories in a fixed AdS$_5$ background.  We apply our formalism to vector multiplets in Sect.\ IV and to hypermultiplets in Sect.\ V.  We conclude with some brief remarks in Sect.\ VI.

\section{Nonlinear Sigma Model in Mink$_5$}

Four-dimensional flat-space $N=2$ supersymmetric sigma models were constructed long ago using both component fields \cite{BaggerWitten,deWitVandoren} and superfields \cite{HuKLR,Harmonic,Projective}.  Supersymmetry requires that the complex scalar fields be the coordinates of a hyper-K\"ahler manifold.

Such theories are relevant to five dimensions because five-dimensional spinors split naturally into two four-dimensional spinors.  For that reason, five-dimensional supersymmetric theories can be described in terms of $N=1$ superfields:  the single five-dimensional supersymmetry splits into two four-dimensional supersymmetries.  The first is manifest in superspace, while the second takes one superfield into another.  The full five-dimensional structure emerges in the closure of the first and second supersymmetries: by Lorentz invariance, they close into a translation along the fifth dimension.  

In previous work \cite{BX1}, we extended \cite{HuKLR} to five-dimensional Minkowski space.  Invariance of the action and closure of algebra require the lowest components of the chiral superfields to be the coordinates of a hyper-K\"ahler manifold, just like $N=2$ in four dimensions \cite{Townsend}.  

To see this explicitly, consider the most general (two-derivative) action for $2n$ chiral superfields in flat space.  We use $N=1$ superspace, so the action is
\bea  \label{eq:5d}
S &=&
 \int\!d^5x d^4\theta \, K ( \Phi^b, \bar{\Phi}^{c*} ) \\
&&+\ \left[ \int\!d^5x d^2\theta  \, [H_a(\Phi^b) \partial_5 \Phi^a + P(\Phi^b)]
+ {\rm h.c.}\right] . \nn
\eea
Here $K$ is the K\"ahler potential, a real function of the superfields, and $H_a$ and $P$ are holomorphic.  By construction, the first supersymmetry is manifest.  The second is given by
\be \label{eq:2nd}
\delta_{\eta} \Phi^{a} \ =\ \frac{1}{2}  J^{ab} \bar{D^2} [ K_b (\theta \eta +
\bar{\theta} \bar{\eta} ) ] - 4  J^{ab} P_b\, \theta \eta,
\ee
where  $J_{ab}= H_{a,b} - H_{b,a}$ is an antisymmetric, holomorphic, covariantly constant tensor that satisfies $J^{ac}  J_{cb} = \delta^{a}{}_{b}$.  (The existence of such a tensor is what makes a K\"ahler manifold hyper-K\"ahler.)  Note that this formalism obscures the five-dimensional Lorentz invariance.  It becomes evident after expanding in terms of component fields and integrating out the auxiliary fields through their equations of motion. (For details, see \cite{BX1}.)

When $P=0$, it is not hard to show that the\break action (\ref{eq:5d}) is invariant under the second supersymmetry transformation.  The transformations close into the five-dimensional algebra with the help of the superfield equations of motion.  When $P \ne 0$, closure imposes  conditions on $X^a = i  J^{ab} P_b$:
\bea  \label{eq:killing}
\nabla_{a} \bar{X}_{\bar{b}} + \bar{\nabla}_{\bar{b}} X_a &=&0\nn\\
\label{eq:tri-holo}
  J^{a}{}_{\bar{b}} \, \bar{\nabla}_{\bar{e}} \bar{X}^{\bar{b}}
-  J^{c}{}_{\bar{e}} \, \nabla_{c} X^a &=& 0.
\eea
The first implies that $X^a$ is a Killing vector.  The second says it must be tri-holomorphic, in which case the isometry leaves the three complex structures invariant.

\section{Warped Superspace}

In this section we construct a modified four-dimensional $N=1$ superspace that is appropriate for five-dimensional supersymmetric theories in an AdS$_5$ background.  We take our bosonic coordinates to be $x^n$ and $z = x^5$, where $n=1,..., 4$, with (warped) metric
\be
\label{metric}
ds^2 \ =\  e^{-2 \lambda z} \eta_{mn} dx^m dx^n +  dz^2
\ee
and cosmological constant $\lambda$.  The metric admits 15 isometries that obey the SO(4,2) algebra.  The AdS$_5$ superalgebra is SU(2,2$|$1), which has the following commutation relations \cite{Zumino, vanProeyen, BKV, Clark}:
\bea 
\{{\mathbb Q}_A, \bar{\mathbb Q}_B \} &=& 2\Gamma^M_{AB} \hat{P}_M + 2i \lambda \Sigma^{MN}_{AB} \hat{M}_{MN}+6 \lambda \delta_{AB}R \nn\\[1mm]
[\hat{M}_{MN}, \hat{M}_{LR}] &=& 2i (\eta_{[ML}\hat{M}_{N]R}-\eta_{[MR}\hat{M}_{N]L}) \nn \\[0mm]
[\hat{M}_{MN}, \hat{P}_{L}] &=& 2i \eta_{[ML}\hat{P}_{N]} \nn \\[0mm]
[\hat{P}_{M}, \hat{P}_N] &=& -i\lambda^2 \hat{M}_{MN}  \nn \\[0mm]
[\hat{M}_{MN}, {\mathbb Q}_A] &=& -\frac{1}{2}(\Sigma_{MN}{\mathbb Q})_A  \nn \\[0mm]
[\hat{M}_{MN}, \bar{\mathbb Q}_A]&=& +\frac{1}{2}(\bar{\mathbb Q}\Sigma_{MN})_A  \nn\\[1mm]
[\hat{P}_{M}, {\mathbb Q}_A] &=& +\frac{\lambda}{2}(\Gamma_M {\mathbb Q})_A   \nn \\[0mm]
[\hat{P}_{M}, \bar{\mathbb Q}_A] &=& -\frac{\lambda}{2}(\bar{\mathbb Q}\Gamma_M )_A  \nn \\[0mm]
[R, {\mathbb Q}_A] &=& +\frac{1}{2}{\mathbb Q}_A   \nn \\[0mm]
[R, \bar{\mathbb Q}_A] &=& -\frac{1}{2}\bar{\mathbb Q}_A  . 
\label{ads5}
\eea
The bosonic generators $\hat{M}_{MN}$ and $\hat{P}_{M}$ describe the 15 isometries.  Writing the generators ${\mathbb Q}_A$, $\hat{M}_{m5}$ and $\hat{P}_m $ in four-dimensional form, 
\bea 
&&\hat{P}_m = \frac{1}{2} (P_m + K_m), \quad  \lambda \hat{M}_{m5} = -\frac{1}{2} (P_m - K_m) \nn \\
&& \mathbb{Q}_A = (Q_{\alpha}, i \bar{S}^{\dot{\alpha}})^T,
\eea
and dropping the hats, we obtain a realization of SU(2,2$|$1) in a four-dimensional notation that is appropriate for our construction,
\bea \label{cft4}
\{Q_\alpha, \bar{Q}_{\dot\alpha}\} &=& 2\sigma^a_{\alpha\dot\alpha}P_a, \quad
\{S_\alpha,\bar{S}_{\dot\alpha}\} =  2\sigma^a_{\alpha\dot\alpha} K_a \nn \\[0mm]
\{Q_\alpha,S_\beta \} &=& 2 \epsilon_{\alpha\beta}P_5 + i 6 \lambda \epsilon_{\alpha\beta}R - 2 \lambda \sigma^{ab}_{\alpha\beta}M_{ab} \nn \\[3mm]
[K^a, Q_\alpha] &=& i\lambda \sigma^a_{\alpha\dot\alpha} \bar{S}^{\dot\alpha}, \quad [K^a, \bar{Q}_{\dot\alpha}]= i\lambda S^\alpha \sigma^a_{\alpha\dot\alpha} \nn \\[0mm]
[P_a, S_\alpha ] &=& i\lambda \sigma_{a\alpha\dot\alpha}\bar{Q}^{\dot\alpha}, \quad [P_a, \bar{S}_{\dot\alpha}]= i \lambda Q^\alpha \sigma_{a\alpha\dot\alpha} \nn \\[3mm]
[P_5, P_a] &=& -i\lambda P_a, \quad [P_5, K_a]=+i \lambda K_a \nn \\[3mm]
[R, Q_\alpha] &=& +\frac{1}{2}Q_\alpha, \quad\ [R, \bar{Q}_{\dot\alpha}] = -\frac{1}{2}\bar{Q}_{\dot\alpha} \nn \\[0mm]
[P_5, Q_\alpha] &=& -i\frac{\lambda}{2}Q_\alpha, \quad [P_5, \bar{Q}_{\dot\alpha}]= -i\frac{\lambda}{2}\bar{Q}_{\dot\alpha} \nn \\[0mm]
[R,S_\alpha ] &=& -\frac{1}{2}S_\alpha, \quad\ \ [R,\bar{S}_{\dot\alpha}] = +\frac{1}{2}\bar{S}_{\dot\alpha} \nn \\[0mm]
[P_5,S_\alpha] &=& +i\frac{\lambda}{2}S_\alpha, \quad\ [P_5,\bar{S}_{\dot\alpha}]=+i\frac{\lambda}{2}\bar{S}_{\dot\alpha}.
\eea
The four-dimensional SO(3,1) generators $M_{mn}$ have the usual commutation relations with the other generators and with themselves.  (See \cite{Clark} for a slightly different notation.)  In the limit of $\lambda \rightarrow 0$, eqs.~(\ref{ads5}) and (\ref{cft4}) reduce to the ordinary $N=2$ super-Poincar\'e algebra in five flat spacetime dimensions.

It is easy to see that the generators $P_m$, $P_5$, $Q_{\alpha}$ and $M_{mn}$ form a subalgebra of SU(2,2$|$1).  In what follows we use this subalgebra to construct ``warped superspace"  -- a version of $N=1$ superspace compatible with the AdS$_5$ algebra.  (Full AdS$_5$ $N=2$ superspace requires a different coset.  The results obtained here can also be found using that approach \cite{5dss,diffcoset}.)  We start by forming the coset element $\Omega$:
\be
\Omega \equiv e^{ i(\theta Q + \bar{\theta} \bar{Q})}\, e^{ iP_m x^m} e^{  i P_5 z}.
\ee
Symmetry transformations $g$ act on $\Omega$ by left multiplication,
\be
\Omega \rightarrow \Omega' = g \Omega.
\ee
This induces a transformation $(x, z, \theta, \bar\theta) \rightarrow (x', z', \theta', \bar\theta')$ on superspace.

The superfield $\Phi (x, z, \theta, \bar\theta)$ is a function on superspace defined by
\be
\Phi (x, z, \theta, {\bar\theta}) \equiv  \Omega\, \Phi(0),
\ee
where $\Phi(0)$ transforms in a finite-dimensional representation of SO(3,1).  By construction, the superfield transforms covariantly under $g$.

Given $\Omega$, we compute the left and right vielbeins $E^{L}_M{}^A$ and $E^{R}_M{}^A$ in terms of the left and right Maurer-Cartan forms, 
\be
iP_A \, E^{L}_M{}^A = \partial_M  \Omega \, \Omega^{-1}, \quad
iP_A \, E^{R}_M{}^A =  \Omega^{-1} \partial_M  \Omega.
\ee
Here we use a compact notation in which $A = (a,5,\alpha,\dot\alpha)$, and likewise for $M$.  We use the inverse vielbeins to construct left and right differential operators $D^L_A = E^{L}_A{}^M \partial_M$ and $D^R_A = E^{R}_A{}^M \partial_M$.  It is an exercise to show that the right derivatives close into the original algebra, while the left derivatives close into the algebra with the opposite sign on the right-hand side.  (For a detailed explanation, see \cite{Butter}.)

The operators $D^L_A$ and $D^L_A$ act on the coordinates of (scalar) superfields $\Phi (x, z, \theta, \bar\theta)$.  The $D^L_A$ generate isometries,
\be
\delta \Phi(x, z, \theta, \bar\theta) \equiv  \xi^A D^L_A \Phi(x, z, \theta, \bar\theta) =   i \xi^A P_A \Omega\, \Phi(0),
\ee
while the $D^R_A$ are covariant derivatives,
\be
{\cal D}_A \Phi(x, z, \theta, \bar\theta) \equiv D^R_A \Phi(x, z, \theta, \bar\theta) =  i \Omega P_A \Phi(0).
\ee
The right vielbein serves as the (invariant) vielbein of the superspace.

For the case at hand, it is easy to compute the covariant derivatives.  They are given by
\bea
{\cal D}_a = D^R_a &=& e^{\lambda z} \partial_a\nn\\[1mm]
{\cal D}_5 = D^R_5 &=& \partial_5 \nn \\[1mm]
{\cal D}_\alpha = D^R_\alpha &=& e^{\frac{1}{2}\lambda z} D_{\alpha} \nn\\[1mm]
\bar{{\cal D}}_{\dot\alpha} = {\bar D}^R_{\dot\alpha} &=& e^{\frac{1}{2}\lambda z} {\bar D}_{\dot\alpha},
\eea
where $D_{\alpha}$ and ${\bar D}_{\dot\alpha}$ are the spinor derivatives in ordinary $N=1$ superspace,
\bea
D_{\alpha} &=& \frac{\partial}{\partial\theta^\alpha} + i \sigma^m_{\alpha\dot\alpha}{\bar\theta}^{\dot\alpha}\partial_m
\nn\\
{\bar D}_{\dot\alpha} &=& -\frac{\partial}{\partial{\bar\theta}^{\dot\alpha}} - i \theta^\alpha \sigma^m_{\alpha\dot\alpha}\partial_m.
\eea
The spinor covariant derivatives take a simple form in terms of the curly variables $\vartheta = e^{-\frac{1}{2}\lambda z} 
\theta$,
\bea
{\cal D}_{\alpha} &=& \frac{\partial}{\partial\vartheta^\alpha} + i \sigma^a_{\alpha\dot\alpha}{\bar\vartheta}^{\dot\alpha}e^m{}_a \partial_m
\nn\\
{\bar {\cal D}}_{\dot\alpha} &=& -\frac{\partial}{\partial{\bar\vartheta}^{\dot\alpha}} - i \vartheta^\alpha \sigma^a_{\alpha\dot\alpha} e^m{}_a \partial_m.
\eea
The bosonic part of the vielbein is $e_m{}^a = e^{-\lambda z} \delta_m{}^a$ and $e_5{}^5=1$, consistent with the metric (\ref{metric}).  (The curly $\vartheta$ are essentially the new $\Theta$ variables of \cite{BW}.)

In a similar fashion, one can compute the differential operators that generate the isometries.  They are
\bea
\label{isom}
i{\cal P}_a  = D^L_a &=&  \partial_a \nn \\
i{\cal P}_5 = D^L_5 &=&  \partial_5 + \lambda x^m \partial_m  + \frac{\lambda}{2} \left(\theta^{\alpha}\frac{\partial}{\partial \theta^{\alpha}} +\bar{\theta}_{\dot{\alpha}} \frac{\partial}{\partial \bar{\theta}_{\dot\alpha}} \right) \nn \\
{\cal Q}_\alpha = D^L_\alpha &=& Q_{\alpha} \nn\\[2mm]
\bar{{\cal Q}}_{\dot\alpha}  = {\bar D}^L_{\dot\alpha} &= &{\bar Q}_{\dot\alpha},
\eea
where 
\bea
Q_{\alpha} &=&  \frac{\partial}{\partial\theta^\alpha} - i \sigma^m_{\alpha\dot\alpha}{\bar\theta}^{\dot\alpha}\partial_m
\nn\\
{\bar Q}_{\dot\alpha} &=& -\frac{\partial}{\partial{\bar\theta}^{\dot\alpha}} + i \theta^\alpha \sigma^m_{\alpha\dot\alpha}\partial_m
\eea
are the usual $N=1$ supersymmetry generators in flat superspace.  In terms of the curly $\vartheta$ variables, they are
\bea
\label{curlyQ}
Q_{\alpha} &=& e^{-\frac{1}{2}\lambda z}\left(\frac{\partial}{\partial\vartheta^\alpha} - i \sigma^a_{\alpha\dot\alpha}{\bar\vartheta}^{\dot\alpha}e^m{}_a \partial_m\right)
\nn\\
{\bar {Q}}_{\dot\alpha} &=& e^{-\frac{1}{2}\lambda z}\left(-\frac{\partial}{\partial{\bar\vartheta}^{\dot\alpha}} + i \vartheta^\alpha \sigma^a_{\alpha\dot\alpha} e^m{}_a \partial_m\right),
\eea
as well as 
\be
 i{\cal P}_5 = \partial_5 + \lambda x^m \partial_m + \frac{\lambda}{2} \left(\vartheta^{\alpha}\frac{\partial}{\partial \vartheta^{\alpha}} +\bar{\vartheta}_{\dot{\alpha}} \frac{\partial}{\partial \bar{\vartheta}_{\dot\alpha}} \right) .
\ee
These differential operators are the essential ingredients of warped superspace.  By construction, the $D^L_A$ commute with the $D^R_A$.  The $D^R_A$ realize the supersymmetry algebra, while the $D^L_A$ realize the algebra with the opposite sign.

The warp factor in (\ref{curlyQ}) is very intriguing.  It suggests that we take $\xi_K \equiv e^{-\frac{1}{2}\lambda z} \xi$ as half a Killing spinor on AdS$_5$.  Indeed, the Killing equation on AdS$_5$ is as follows: 
\be
D_M \Xi_K^i - i\frac{\lambda}{2} \Gamma_M (\tau_3)^i{}_j \Xi_K^j = 0.
\ee
Its solution is
\be
\label{Killingspinor}
\Xi^1_K = 
\begin{pmatrix}
\xi_K - i \lambda \, \sigma^a \bar\eta_K\, e^a{}_m \, x^m \\ \bar\eta_K 
\end{pmatrix},
\ee
and likewise for $\Xi^2_K$,
where $\xi_K = e^{-\frac{1}{2}\lambda z} \xi$ and $\eta_K = e^{+\frac{1}{2}\lambda z} \eta$.

We now have what we need to write supersymmetric AdS$_5$ actions.  We start by defining warped chiral superfields in complete analogy to $N=1$:
\be
{\bar{\cal D}}_{\dot\alpha} \Phi = 0.
\ee
This condition is consistent with integrability since
\be
[{\cal D}_5, \bar{{\cal D}}_{\dot\alpha}]= \frac{\lambda}{2}\, \bar{{\cal D}}_{\dot\alpha},
\quad
[{\cal D}_m, \bar{{\cal D}}_{\dot\alpha}]=0.
\ee
Then, using these superfields, it is easy to write the most general two-derivative action, in analogy with (\ref{eq:5d}):
\bea  \label{eq:5dwarped}
S &=&
 \int\!d^5x d^4\vartheta\, \det e \, K ( \Phi^b, \bar{\Phi}^{c*} )\\
&&+ \left[  \int\!d^5x d^2\vartheta \,\det e \,  [H_a(\Phi^b) {\cal D}_5 \Phi^a + P(\Phi^b)]
+ {\rm h.c.}\right] . \nn
\eea
Because ${\cal Q}_\alpha \sim {\cal D}_\alpha$ up to an $x^m$ derivative, this action is invariant under the isometries generated by ${ \cal Q}_\alpha$ (and likewise for ${\bar{\cal Q}}_{\dot\alpha})$.  By construction, it is also invariant under the isometries generated by ${\cal P}_m$ and ${\cal P}_5$.  In terms of $N=1$ superspace, (\ref{eq:5dwarped}) reduces to
\bea  \label{eq:5dwarped2}
S &=&
 \int\!d^5x d^4\theta\, e^{-2 \lambda z}\, K ( \Phi^b, \bar{\Phi}^{c*} ) \\
&&+ \left[  \int\!d^5x d^2\theta \,e^{-3 \lambda z}\, [H_a(\Phi^b) \partial_5 \Phi^a + P(\Phi^b)]
+ {\rm h.c.}\right] . \nn
\eea

Component fields and $\theta$ expansions are defined using the covariant derivatives.  Indeed, as in flat space, we define
\be
\Phi | = A, \quad {\cal D}_\alpha \Phi | = \sqrt{2} \psi_\alpha, \quad {\cal D}^2 \Phi | = -4 F.
\ee
This gives rise to the following $\theta$ expansion
\bea 
\Phi &=&  A + \sqrt{2}\,\vartheta\psi + \vartheta\vartheta F + i\vartheta \sigma^a \bar\vartheta e^m{}_a \partial_m A +\ldots \\
&=& A + \sqrt{2}\,e^{-\frac{1}{2} \lambda z} \theta\psi + e^{-\lambda z}\,\theta\theta F + i\theta \sigma^m \bar\theta \partial_m A + \ldots \nn
\eea
where $\vartheta = e^{-\frac{1}{2} \lambda z} \theta$ and $e^m{}_a = e^{\lambda z} \delta^m{}_a$.  The ${\cal D}$ provide just the right warp factors to give the correct component results \cite{MP}.  Indeed, the component supersymmetry transformations are
\bea
\delta A &\equiv& \delta \Phi | = (\xi^\alpha {\cal Q}_\alpha + {\bar\xi}_{\dot\alpha} {\bar{\cal Q}}^{\dot\alpha})
\Phi | =  \sqrt{2}\, \xi_K^\alpha \psi_\alpha \nn\\[2mm]
\delta \psi_\alpha & \equiv & \frac{1}{\sqrt 2} \delta {\cal D}_\alpha \Phi | = \frac{1}{\sqrt 2} {\cal D}_\alpha (\xi^\alpha {\cal Q}_\alpha + {\bar\xi}_{\dot\alpha} {\bar{\cal Q}}^{\dot\alpha}) \Phi | \nn \\[0.5mm]
&= & i \sqrt{2}\, \sigma^a_{\alpha\dot\alpha} {\bar\xi}_K^{\dot\alpha}\, e^m{}_a \partial_m A +\sqrt{2}\, \xi_{K\alpha} F.
\eea
These are half the component supersymmetry transformations in AdS$_5$, recognizing that $\xi_K = e^{-\frac{1}{2} \lambda z} \xi_\alpha$ is half of the AdS$_5$ Killing spinor.

\section{Vector Multiplet in AdS$_5$}

We are now in position to construct rigidly supersymmetric theories in an AdS$_5$ background.  We first consider the five-dimensional vector multiplet.  To set notation, we start by writing the action and supersymmetry transformations in flat space.

The five-dimensional vector multiplet contains a five-dimensional vector gauge field $A_{M}$, a four-component Dirac gaugino $\lambda_i$, and a real scalar $\Sigma$.  They can be collected into two $N=1$ superfields \cite{AGW, MP}:  a vector superfield $V$ and a chiral superfield $\chi$,
\begin{eqnarray}
V &=& -\theta \sigma^m \tBar A_{m}
+ i \tBar^2 \theta \lambda_1
-  i \theta^2 \tBar \bar{\lambda}_1
+\half\tBar^2 \theta^2 D \nonumber \\
\chi &=& \frac{1}{2}\left(\Sigma + i A_5\right)
+ \sqrt{2}\theta \lambda_2 + \theta^2 F.
\end{eqnarray}
In this expression, $V$ is in the Wess-Zumino gauge and $\chi$ is in the chiral basis. The action is given by
\bea
S &=& \frac{1}{4 g^2} \bigg[ \int\!d^5x d^2\theta \, W^{\alpha} W_{\alpha} + {\rm h.c.} \bigg]\nn \\[1mm]
&&\ +\frac{1}{g^2} \int\!d^5x d^4\theta \, (\chi +\bar{\chi} - \partial_5 V )^2  .
\eea
It is invariant under the following gauge transformations:
\bea
\delta V &=& \Lambda + \bar\Lambda \nn\\
\delta \chi &=& \partial_5 \Lambda,
\eea
where $\Lambda$ is a chiral superfield.  The action describes five-dimensional supersymmetric Yang-Mills theory, with a U(1) gauge group, written in $N=1$ superspace.  (The action and transformations can be readily generalized to nonabelian gauge groups.)

With this construction, the first supersymmetry is manifest.  The second is a little trickier.  It rotates the two superfields into each other:
\begin{eqnarray} \label{eq:2nda}
\delta_{\eta} V &=& 2 ( \chi + \bar{\chi} -  \partial_5 V )
( \theta \eta + \bar{\theta} \bar{\eta} )\nonumber \\[1mm]
\delta_{\eta} \chi &=& -  \eta  W .
\end{eqnarray}
A small calculation shows that these transformations close into the five-dimensional flat-space supersymmetry algebra.
 
Let us now consider an AdS$_5$ background.  A natural guess is to replace the flat-space $D$'s with their curly AdS$_5$ counterparts.  The field strength is 
\be
{\cal W}_{\alpha} = -\frac{1}{4} {\bar{\cal D}}^2 {\cal D}_{\alpha} V;
\ee
it is still chiral because of the $\cal D$ algebra.  The action is
\bea\label{eq:5Dwarped} 
S &=& \frac{1}{4 g^2} \bigg[ \int\!d^5x d^2\vartheta \, \det e\, {\cal W}^{\alpha} {\cal W}_{\alpha} + {\rm h.c.} \bigg] \nn \\[1mm]
&&+\frac{1}{g^2} \int\!d^5x d^4\vartheta \, \det e\, (\chi +\bar{\chi} - {\cal D}_5 V )^2  .
\eea
As above, it is invariant under the following gauge transformations:
\bea
\delta V &=& \Lambda + \bar\Lambda \nn\\[1mm]
\delta \chi &=& {\cal D}_5 \Lambda.
\eea
The superfield $\chi$ remains chiral because of the ${\cal D}$ algebra.

The action (\ref{eq:5Dwarped}) can be easily written in terms of ordinary $N=1$ superspace.  One finds
\bea\label{ads_v}
S &=& \frac{1}{4 g^2} \bigg[ \int\!d^5x d^2\theta \, {W}^{\alpha} {W}_{\alpha} + {\rm h.c.} \bigg] \nn \\[1mm]
&&\ +\frac{1}{g^2} \int\!d^5x d^4\theta \, e^{-2\lambda z} (\chi +\bar{\chi} - \partial_5 V )^2  .
\eea
where ${W}_{\alpha} = -\frac{1}{4} {\bar{D}}^2 {D}_{\alpha} V$.  This is the same action as in \cite{MP}.

Let us now determine the supersymmetry transformation laws for (\ref{eq:5Dwarped}).  The first supersymmetry is manifest.  A plausible guess for the second supersymmetry is
\bea \label{ads2nd}
\nonumber
\delta_{\eta} V &=& 2 ( \chi + \bar{\chi} -  {\cal D}_5 V )
( \vartheta \eta_K + \bar{\vartheta} \bar{\eta}_K) - \lambda\, \tilde{\eta}_s^A {\cal D}_A V   \\[1mm]
\delta_{\eta} \chi &=& - \eta_K\, {\cal W} -  \lambda \,\tilde{\eta}_s^A {\cal D}_A \chi ,
\eea
where $ \tilde{\eta}_s^A  {\cal D}_A = \tilde{\eta}_s^a {\cal D}_a + \tilde{\eta}_s^{\alpha}  {\cal D}_{\alpha} + \bar{\tilde{\eta}}_{s\dot{\alpha}} \bar{{\cal D}}^{\dot{\alpha}} $ and $ \tilde{\eta}_s^A $ is the four-dimensional S-supersymmetry transformation \cite{BuchKuzenko} in warped superspace:
\bea
\nn
\tilde{\eta}_s^a  &=& - 2(\vartheta \sigma^a \bar{\sigma}^b \eta_K + \bar{\eta}_K \bar{\sigma}^b \sigma^a 
\bar{\vartheta}) x^m e_{mb} \\[1mm] \nn 
&& +\ 2i ( \bar{\vartheta}^2 \vartheta \sigma^a \bar{\eta}_K + \vartheta^2 \bar{\vartheta} \bar{\sigma}^a \eta_K ) \\
\tilde{\eta}_s^\alpha &=&  - \frac{i}{8} ({\bar{\cal D}} {\bar\sigma}_a)^\alpha  \tilde{\eta}_s^a , 
\eea
where $\eta_K = e^{+\frac{1}{2}\lambda z} \eta$ and $e_{mb} = e_m{}^{a} \eta_{ab}$.  In flat $N=1$ superspace, (\ref{ads2nd}) reduces to
\bea 
\nonumber
\delta_{\eta} V &=& 2 ( \chi + \bar{\chi} -  \partial_5 V )
( \theta \eta + \bar{\theta} \bar{\eta}) - \lambda\, \eta_s^A D_A V   \\
\delta_{\eta} \chi &=& - e^{2 \lambda z} \eta \, W -  \lambda \,\eta_s^A D_A \chi ,
\eea
where $ \eta_s^A $ is the four-dimensional S-supersymmetry transformation in flat spacetime \cite{BuchKuzenko},
\bea
\nn
\eta_s^a  &=& - 2(\theta \sigma^a \bar{\sigma}^b \eta + \bar{\eta} \bar{\sigma}^b \sigma^a 
\bar{\theta}) x^m \hat{e}_{mb} \\[1mm] 
&& +\ 2i ( \bar{\theta}^2 \theta \sigma^a \bar{\eta} + \theta^2 \bar{\theta} \bar{\sigma}^a \eta )\nn\\
{\eta}_s^\alpha &=&  - \frac{i}{8} ({\bar{D}} {\bar\sigma}_a)^\alpha  {\eta}_s^a , 
\eea
The parameters $\tilde{\eta}_s^A$ and $ \eta_s^A $ are related as follows,
\be \label{warpflat}
 \tilde{\eta}_s^A = (\tilde{\eta}_s^a, \tilde{\eta}_s^{\alpha}, \bar{\tilde{\eta}}_{s\dot{\alpha}} ) 
 = ( e^{-\lambda z} \eta_s^a, e^{- \frac{1}{2} \lambda z} \eta_s^{\alpha}, e^{- \frac{1}{2} \lambda z} \bar{\eta}_{s\dot{\alpha}} ). \\
\ee

We have checked that the action (\ref{eq:5Dwarped}) is invariant under the transformation (\ref{ads2nd}). When $\lambda$ goes to zero, (\ref{ads2nd}) reduces to (\ref{eq:2nda}), the supersymmetry transformation in flat spacetime.  In fact, comparing (\ref{ads2nd}) with (\ref{Killingspinor}), a consistent story emerges: the second AdS$_5$ supersymmetry is the combination of a second flat-space supersymmetry (with parameter $\eta_K$) and a four-dimensional superconformal supersymmetry with the same parameter.

The closure of two second supersymmetry transformations (\ref{ads2nd}) produces a transformation of the form,
\bea \label{ads_boost}
\nonumber
\delta_{f} V &=&  f_K^a {\cal D}_a V + 2 \lambda x^a f_{Ka} (\chi + \bar{\chi} - {\cal D}_5 V ) 
+  \lambda^2\,  \tilde{\xi}_f^A {\cal D}_A V   \\ 
\delta_{f} \chi &=& f_K^a {\cal D}_a \chi + \frac{i}{4} \lambda  f_{Ka} 
{ \bar{\cal D}}^2 ({\cal D}^{\alpha} V \sigma^a_{\alpha \dot{\alpha}} \bar{\vartheta}^{\dot{\alpha}} )
 + \lambda^2\,  \tilde{\xi}_f^A {\cal D}_A \chi ,\nn\\
\eea
where $f_K^a = e^{+\lambda z} f^a $, and $\tilde{\xi}_f^A $ is the four-dimensional superconformal Killing vector of special conformal transformations in warped superspace.  It is related to the flat one \cite{BuchKuzenko} in a similar way as (\ref{warpflat})
\bea
\nonumber
\tilde{\xi}_f^a  &=&  f_K^a x^2 -2 x^a (x \cdot f_K) + 2 \epsilon^{abcd} f_{Kb} x_c  \vartheta {\sigma}_d \bar{\vartheta}
-  \vartheta^2  \bar{\vartheta}^2 f_K^a \\ 
\tilde{\xi}_f^\alpha &=&  - \frac{i}{8} ({\bar{\cal D}} {\bar\sigma}_a)^\alpha  \xi_f^a .
\eea 
It is straightforward to check that (\ref{ads_boost}) is an isometry of the AdS$_5$ background, consistent with (\ref{cft4}), written in terms of superfield transformations.

\section{Nonlinear Sigma Model in AdS$_5$}

We now apply what we have learned to construct the nonlinear sigma model that describes hypermultiplets in an AdS$_5$ background.  (For harmonic and projective\break superspace, see \cite{5dss}.)  In warped superspace, we write the action as 
\bea \label{ads5H_w}
S &=&
 \int\!d^5x d^4\vartheta \,\det e\, K ( \Phi^b, \bar{\Phi}^{c*} ) \\
&&+\ \left[ \int\!d^5x d^2\vartheta  \,\det e\, [H_a(\Phi^b) {\cal D}_5 \Phi^a + P(\Phi^b)]
+ {\rm h.c.}\right] , \nn
\eea
which, in flat superspace, becomes
\bea 
S &=&
 \int\!d^5x d^4\theta \,e^{-2 \lambda z} K ( \Phi^b, \bar{\Phi}^{c*} ) \\
&&+\ \left[ \int\!d^5x d^2\theta  \,e^{-3 \lambda z} [H_a(\Phi^b) \partial_5 \Phi^a + P(\Phi^b)]
+ {\rm h.c.}\right] . \nn
\eea
In the flat spacetime limit, where $\lambda \rightarrow 0$, the gradient of the superpotential $P$ must be a tri-holomorphic Killing vector $Z^a = i J^{ab} P_b$ \cite{BX1}. When $\lambda \neq 0$, one can redefine the holomorphic one-form $H_a $,
\be \label{shiftH}
H_a \rightarrow H_a - \frac{1}{3\lambda}P_a,
\ee
and integrate by parts to absorb the superpotential $P$ into $H_a$.  Therefore, without loss of generality, we can take the action to be
\bea \label{ads5H_f}
S &=&
 \int\!d^5x d^4\vartheta \,\det e\, K ( \Phi^b, \bar{\Phi}^{c*} ) \nn\\
&&\ +\ \left[ \int\!d^5x d^2\vartheta  \,\det e \,H_a(\Phi^b) {\cal D}_5 \Phi^a
+ {\rm h.c.}\right]  \nn\\[2mm]
&=&
 \int\!d^5x d^4\theta \,e^{-2 \lambda z} K ( \Phi^b, \bar{\Phi}^{c*} ) \\
&&\ +\ \left[ \int\!d^5x d^2\theta  \,e^{-3 \lambda z} \,H_a(\Phi^b) \partial_5 \Phi^a
+ {\rm h.c.}\right] . \nn
\eea
As we will see, $H_a$ obeys a constraint whose solution is determined up to a tri-holomorphic Killing vector, which precisely reflects the degree of freedom in (\ref{shiftH}).

We take the following ansatz for the second supersymmetry transformation,
\bea \label{5Dlaw}
\delta_{\eta} \Phi^a &=& \frac{1}{2} \bar{{\cal D}}^2 [  J^{ab} K_b ( \vartheta \eta_K + 
\bar{\vartheta} \bar{\eta}_K ) ] \nn \\
&&\ +\ 12 \lambda  J^{ab} H_b \vartheta \eta_K\ -\ \lambda {\tilde\eta_s}^A {\cal D}_A \Phi^a \nn
\\[2mm]
&=& \frac{1}{2} e^{\lambda z} \bar{D}^2 [  J^{ab} K_b ( \theta \eta + 
\bar{\theta} \bar{\eta} ) ] \nn\\
&&\ + \ 12 \lambda  J^{ab} H_b \theta \eta\ - \ \lambda \eta_s^A D_A \Phi^a ,
\eea
where ${\tilde\eta}^A_s$ and $\eta^A_s$ are the four-dimensional superconformal Killing supervectors for the S-supersymmetry.  The transformation (\ref{5Dlaw}) preserves the chirality of $\Phi^a$.  In the limit $\lambda \rightarrow 0$, the action (\ref{ads5H_w}) and the transformation (\ref{5Dlaw}) reduce to the correct flat space forms.

The invariance of the action (\ref{ads5H_f}) under (\ref{5Dlaw}) imposes a constraint on the holomorphic one-form $H_a$:
\be \label{AdSKP}
K_a  J^{ab} H_b -iF = K_{\bar{a}} \bar{ J}^{\bar{a}\bar{b}} \bar{H}_{\bar{b}} + i \bar{F} \equiv D_H.
\ee  
Here $F$ is a holomorphic function and $D_H$ is real.  Equation (\ref{AdSKP}) represents a new restriction on the hyper-K\"ahler target space in an AdS$_5$ background -- one that does not appear in the flat space limit.

This constraint (\ref{AdSKP}) is also necessary to ensure the five-dimensional invariance of the action.  To see this, we compute the component Lagrangian and integrate out the auxiliary fields $F^a$.  This gives rise to the following terms:
\bea \label{LF}
\nn
{\cal L}\ &\subseteq& - \det e\, g^{a\bar{b}}\, ( J_{ab} \bar{ J}_{\bar{b}\bar{c}}\, \partial_5 A^b 
\partial_5 A^{*\bar{c}} + 9 \lambda^2 H_a \bar{H}_{\bar{b}}  \\[1mm] 
&&\ - 3 \lambda H_a \bar{ J}_{\bar{b}\bar{c}}\partial_5 A^{*\bar{c}}
- 3 \lambda \bar{H}_{\bar{b}}  J_{ab} \partial_5 A^b ).
\eea
Since $g^{a\bar{b}}  J_{ab} \bar{ J}_{\bar{b}\bar{c}} = g_{b\bar{c}} $, the first term of (\ref{LF}) contributes to the scalar kinetic term.  The second term contributes to the scalar potential.  The cross terms in the second line of (\ref{LF}) are dangerous, however, because they appear to violate the five-dimensional Lorentz symmetry. 

Fortunately, Eq.\ (\ref{AdSKP}) neutralizes the cross terms.  From (\ref{AdSKP}), one finds
\bea \label{pDH}
\nn
 g^{a\bar{b}}H_a  \bar{ J}_{\bar{b}\bar{c}}   & = & \frac{\partial D_H}
{\partial A^{* \bar{c}} } \\
 g^{a\bar{b}} \bar{H}_{\bar{b}}   J_{ac} & = & \frac{\partial D_H}{\partial A^c}  .
\eea
This permits the cross terms to be combined into a total derivative,
\bea
\nn\label{tot}
&&- 3 \lambda g^{a\bar{b}} ( H_a \bar{ J}_{\bar{b}\bar{c}}\partial_5 A^{*\bar{c}} + \bar{H}_{\bar{b}}  J_{ab} \partial_5 A^b ) \\ \nn
&=& -3 \lambda ~(\partial_{\bar{a}} D_H \partial_5 A^{*\bar{a}} + \partial_a D_H \partial_5 A^a )  \\
&=& -3 \lambda ~\partial_5 D_H.
\eea
Substituting (\ref{tot}) into (\ref{LF}), and integrating by parts, we find a second contribution to the AdS$_5$ scalar potential,
\bea \label{V}
\cal{V} &=& 9 \lambda^2 \left( g^{a\bar{b}} H_a \bar{H}_{\bar{b}} - \frac{4}{3} D_H \right) .
\eea
Note that this potential arises even in the absence of a superpotential.

The function $D_H$ is, in fact, a Killing potential.  Its gradient gives rise to a holomorphic Killing vector,
\be 
\label{holokv}
X^a \equiv i J^{ab} H_b = i g^{a\bar b} \frac{\partial D_H}{\partial A^{*\bar b}},
\ee
where
\be\label{KV1}
\nabla_{a} \bar{X}_{\bar{b}} + \nabla_{\bar{b}} X_a =0.
\ee
Since $ H_a = -i  J_{ab} X^b $, the fact that $H_{a,b} - H_{b,a} = J_{ab}$ imposes an additional constraint on $ X^a $:
\be  \label{KV2}
J_{ac} \nabla_b X^c -  J_{bc} \nabla_a X^c =  i J_{ab} ,
\ee
or equivalently,
\be \label{THKVnew}
J^{a}{}_{\bar{b}} \nabla_{\bar{c}} \bar{X}^{\bar{b}}
-  J^{b}{}_{\bar{c}}  \nabla_{b} X^a = -i  J^a{}_{\bar{c}}.
\ee
This modified tri-holomorphic constraint is also needed to close the second supersymmetry transformation (\ref{5Dlaw}). 
It is similar to the condition found in \cite{ButterKuzenko} for ${\cal N} = 2$ hyper-K\"ahler models in AdS$_4$.

It is easy to see that  two different solutions of Eq.\ (\ref{THKVnew}), say $X_1$ and $X_2$, differ by a tri-holomorphic Killing vector, since 
\be
 J^{a}{}_{\bar{b}} \nabla_{\bar{c}} \bar{Z}^{\bar{b}} 
-  J^{b}{}_{\bar{c}}  \nabla_{b} Z^a = 0
\ee
is the tri-holomorphic condition for $Z= X_1-X_2$.  As we saw previously, adding such a vector adds a superpotential to the system.

In general we do not know how to solve the Eqs.\ (\ref{KV1}) and (\ref{KV2}) for $X^a$.  However, they can be readily solved if
\be \label{HKVX}
\nabla_a \hat{X}^b =  \frac{i}{2}~\delta_a{}^{b}.
\ee
Then any solution of (\ref{KV1}) and (\ref{KV2}) can be written as 
\be
X^a = \hat{X}^a + c Z^a,
\ee
where $Z$ is tri-holomorphic and $c$ is a real constant.  Note that the Killing vector $\hat{X}^a$ is holomorphic, but {\it not} tri-holomorphic, in accord with Eq.\ (\ref{THKVnew}). (In fact, $Y  \equiv -2i\hat X$ is a {\it homothetic} Killing vector \cite{sezgin, deWit}. Manifolds admitting such vectors are called hyper-K\"ahler cones or Swann spaces \cite{deWit}.)

Let us now check invariance under ${\cal P}_5$.  Following ({\ref{isom}), we take
the variation of $ \Phi^a $  to be
\be  \label{5DD}
\delta_D \Phi^a =  \Lambda_D \partial_5 \Phi^a + \lambda \epsilon_D^A D_A \Phi^a,
\ee
where $ \epsilon_D^A = (\epsilon_D^a,\ \epsilon_D^{\alpha},\ \bar{\epsilon}_{D\dot{\alpha}} )$ is the four-dimensional superconformal dilatation Killing supervector
\be
\epsilon_D^m = \Lambda_D x^m ,~~  \epsilon_D^{\alpha} = \frac{1}{2} \Lambda_D \theta^{\alpha},
~~ \bar{\epsilon}_{D\dot{\alpha}} = \frac{1}{2} \Lambda_D \bar{\theta}_{\dot{\alpha}}.
\ee
One can check that the action (\ref{ads5H_f}) is invariant under (\ref{5DD}).  This requires that the warped factors be exactly as in (\ref{ads5H_f}), but it does not restrict the K\"ahler potential $K$ or the holomorphic one-form $H_a$. 

In a similar way, we can promote the four-dimensional axial symmetry into the five-dimensional U(1) R-symmetry:
\be \label{5DR}
\delta_R \Phi^a =  \Lambda_R X^a - \epsilon_R^A D_A \Phi^a.
\ee
The first term contains the Killing vector (\ref{holokv}) and the second implements the four-dimensional axial symmetry,
\be
\epsilon_R^m = \Lambda_R \theta \sigma^m \bar{\theta} ,~~  \epsilon_R^{\alpha} = \frac{i}{2} \Lambda_R \theta^{\alpha},
~~ \bar{\epsilon}_{R\dot{\alpha}} =- \frac{i}{2} \Lambda_R \bar{\theta}_{\dot{\alpha}}.
\ee
As in the dilatation case,  invariance of the action (\ref{ads5H_f}) under (\ref{5DR}) imposes no new conditions.

Finally, closing the second supersymmetry (\ref{5Dlaw}) with itself gives the five-dimensional lift of a special conformal transformation,
\bea \label{5DMmu5}
\delta_f \Phi^a &=& e^{2 \lambda z} f^m \partial_m \Phi^a - 2 \lambda y^m f_m \partial_5 \Phi^a \\ \nn
 &-& i \frac{\lambda}{2} e^{\lambda z} \bar{D}^2~[ J^{ab} K_b  \theta \sigma^m \bar{\theta} f_m ] 
 + \lambda^2  \xi^A_f D_A \Phi^a ,
\eea
where $y^m = x^m + i \theta \sigma^m \bar{\theta}$ and $ \xi^A_f$ is four-dimensional\break special conformal Killing supervector.

We close by presenting a simple model, described by two chiral multiplets, $U$ and $V$, with K\"ahler potential $K = U \bar{U} + V \bar{V}$ and antisymmetric tensor $ J_{UV} = - J_{VU}= -J^{UV}=-1 $.  As shown in \cite{deWit}, this hyper-K\"ahler manifold has a rich structure because of its large symmetry group.  First, there is an SU(2) R-symmetry group with Killing vectors  \cite{deWit}
\bea \label{SU2RKV}
\nn
X^3 &=& i U \partial_U + i V \partial_V  \\[1mm]
X^+ &=& -V \partial_{\bar{U}} + U \partial_{\bar{V}} \nn\\[1mm]
X^- &=& -\bar{V} \partial_U + \bar{U} \partial_V  .
\eea
Note that $X^3$ is holomorphic, while $X^+, X^-$ are non-holomorphic, and none of them are tri-holomorphic.  There is also a second SU(2) isometry group, generated by 
\bea \label{SU2TKV} \nn
Z^1 &=& V \partial_U - U \partial_V, \\[1mm]
Z^2 &=& i V \partial_U + i U \partial_V\nn\\[1mm]
Z^3 &=& -i U \partial_U + i V \partial_V  .
\eea
All three of these vectors are tri-holomorphic. 

The AdS$_5$ sigma model corresponding to this hyper-K\"ahler manifold has the action
\bea \label{5Dsimple}
S &=&
\int\!d^5x \,d^4\theta\, e^{-2 \lambda z}\, ( U \bar{U} + V \bar{V} )  \\
&&+ \left[ \int\!d^5x \,d^2\theta \,e^{-3 \lambda z} (H_U \partial_5 U + H_V \partial_5 V)
+ {\rm h.c.}\right] . \nn
\eea
To find the vector $X^a$, we solve $\nabla_a \hat{X}^b = \frac{i}{2} \delta_a{}^b$:
\be
\hat{X}^U = \frac{i}{2}\, U, \quad \hat{X}^V =  \frac{i}{2} \,V . 
\ee
Note that $\hat{X}$ is proportional to the Killing vector $X^3$ in (\ref{SU2RKV}).   We then write $X = \hat{X} + c Z $, with $Z$ tri-holomorphic.  For the case at hand, we take $Z=Z^3$ and compute $X^a$ and $H_a$:
\bea 
\nn
X^U =  i \left(\frac{1}{2} - c \right)  U, &&\qquad X^V =  i \left(\frac{1}{2} +  c\right) V  \\ 
H_U = -\left(\frac{1}{2} +c \right) V, && \qquad H_V = \left(\frac{1}{2} -c\right) U.
\eea
(Choosing $Z^1$ or $Z^2$ leads to the same mass spectrum after a field redefinition.)
The action is then
\bea \nn
S &=& \int d^5x d^4\theta\, e^{-2 \lambda z} \,( U \bar{U} + V \bar{V} ) \\
&&\ +\ \bigg[ \int d^5x d^2\theta e^{-3 \lambda z} \Big[\frac{1}{2} ( U \partial_5 V - V \partial_5 U)\nn\\
&& \  -\ 3 c\lambda U V \Big] + {\rm h.c.} \bigg],
\eea
where the ``superpotential" term emerges after integration by parts.  

To determine the mass spectrum, we calculate the Killing potential according to (\ref{AdSKP}}):
\be
D_H = \frac{1}{2} \big( U \bar{U} + V \bar{V} \big) - c \big(U \bar{U} - V \bar{V}\big).
\ee
Note that $c$ is real, and $F$=0 in (\ref{AdSKP}). The scalar potential is (\ref{V}):
\be
{\cal{V}} =  m^2_{u} U \bar{U} + m^2_{v} V \bar{V},
\ee
where the masses are given by
\bea \label{uvmass}
m^2_{u} &=& \mu^2 - \mu \lambda - \frac{15}{4} \lambda^2 \nn\\
 m^2_{v} &=& \mu^2 + \mu \lambda - \frac{15}{4} \lambda^2,
\eea
with $\mu = -3 \lambda c$, in accord with \cite{Shuster, MP}.  

\section{Discussion and Conclusions}

In this paper we created a warped superspace to\break facilitate the construction of supersymmetric matter couplings in AdS$_5$ backgrounds.  Our superspace is based on ordinary $N=1$ superspace, so it is easy to use and well suited to brane foliations of AdS$_5$.  

We applied our construction to AdS$_5$ vector and\break hyper multiplets.  For vector multiplets, we found results similar to those in flat five-dimensional space.  For\break hypermultiplets, we found that the most general couplings are more restrictive than in flat space.  Not only must the scalar fields parametrize a hyper-K\"ahler manifold, but the manifold itself must be endowed with a holomorphic Killing vector that obeys an inhomogeneous tri-holomorphic condition.  Swann manifolds satisfy the condition, but perhaps others do as well.

While this work was being written up, two papers were posted that considered superspace sigma models in four-dimensional anti de-Sitter backgrounds \cite{Seiberg, ButterKuzenko}.  In our paper we focussed our attention on five-dimensional anti de-Sitter supersymmetry without boundary terms.  It would be interesting to determine the relation between bulk and boundaries for the case at hand \cite{BL}.

\section*{Acknowledgments}

J.B.\ is supported by the US National Science Foundation, grant NSF-PHY-0910467. C.X.\ is supported by the U.S.\ Department of Energy under grant No.\ DE-FG02-97ER41027 and grant No.\ DE-FG02-91ER40681(task B). C.X.\ would like to thank T.E.\ Clark, S.T.\ Love and T.\ ter Veldhuis for thoughtful conversations.  C.X.\ and J.B.\ would also like to thank Jingsheng Li for helpful discussions, and S.\ Kuzenko and G.\ Tartaglino Mazzucchelli for correspondence.

\end{document}